\documentclass[showpacs,amssymb,twocolumn,aps,nofootinbib]{revtex4}
\usepackage{amsmath}
\usepackage{amstext}
\usepackage{amsopn}
\usepackage{amsfonts}
\usepackage{amssymb}
\usepackage{bbm}
\usepackage{accents}
\usepackage{empheq}
\usepackage{graphicx}
\usepackage{epsf}
\usepackage{graphics}
\usepackage{xcolor}
\usepackage[latin1]{inputenc}

\begin{document}

\title{Generalized fluctuation-dissipation theorem in a soluble out of equilibrium model}

\author{A. L. M. Britto$^{a}$, Ashok K. Das$^{b,c}$ and J. Frenkel$^{a}$}
\affiliation{$^{a}$ Instituto de Física, Universidade de São Paulo, 05508-090, São Paulo, SP, Brazil}
\affiliation{$^b$ Department of Physics and Astronomy, University of Rochester, Rochester, NY 14627-0171, USA}
\affiliation{$^c$ Saha Institute of Nuclear Physics, 1/AF Bidhannagar, Calcutta 700064, India}

\begin{abstract}
In the context of an exactly soluble out of equilibrium (quenched) model, we study an extension of the fluctuation-dissipation relation. This involves a modified differential form of this relation, with an effective temperature which may have an explicit dependence on time scales.
\end{abstract}


\maketitle

\section{Introduction}
The fluctuation-dissipation theorem plays an important role in understanding physical systems in thermal equilibrium \cite{callen,kubo,frenkel}. Basically, it says that the statistical fluctuations in a system can be related to the imaginary part of the  response function of the system to a weak external perturbation, through a temperature dependent factor. Historically, this relation has proved quite  useful in the study of several systems, starting with the Brownian motion of a free particle in a liquid \cite{E} as well as the thermal noise in a conductor \cite{J+H}. In a near-equilibrium  situation, it is possible to extend the fluctuation-dissipation theorem through a gradient expansion which is valid only for 
slowly varying processes \cite{G+R,L+M}. The study of quantum field theories out of equilibrium is also of much interest in many branches of physics, like in condensed matter \cite{A+S} or else in the case of heavy ion collision experiments at RHIC. However, when the system is out of equilibrium, such a simple extension of the fluctuation-dissipation relation does not hold in general, since in this regime the non-equilibrium Green's functions are rapidly varying and usually have two independent components (in the thermal average of the two point function). Therefore, it is important, both theoretically and experimentally, to obtain an appropriate generalization in this regime.

The fluctuation-dissipation theorem gives an important relation between the correlated and the retarded Green's functions of a system in thermal equilibrium \cite{callen,kubo,frenkel}. For example, in a scalar field theory if we define the thermal averages
\begin{align}
G_{R} (x,y) & = - i \theta (x^{0}-y^{0}) \langle [\phi(x), \phi (y)]\rangle,\notag\\
G_{c} (x,y) & = -2i C(x,y) = -i \langle [\phi(x), \phi(y)]_{+}\rangle,\label{1}
\end{align}
the fluctuation-dissipation theorem states that their Fourier transforms satisfy ($p_{\mu}$ denotes the conjugate variable to the coordinate difference $x^{\mu}-y^{\mu}$) 
\begin{equation}
C(p) = - \coth \frac{\beta p_{0}}{2}\, \text{Im}\,G_{R} (p).\label{2}
\end{equation}
Namely, the statistical fluctuations in a theory in equilibrium are related to dissipation in the system through a temperature dependent factor. Here $\beta$ is related to the inverse (equilibrium) temperature and we note that the temperature dependent factor in \eqref{2} can also be written as
\begin{equation}
\coth \frac{\beta p_{0}}{2} = 1 + 2 n (p_{0}),\label{3}
\end{equation}
where $n (p_{0}) = 1/(e^{\beta p_{0}} -1)$ denotes the equilibrium Bose-Einstein distribution function.

In equilibrium, the Green's functions depend on the difference of the time coordinates, $t=x^{0}-y^{0}$, because of time translation invariance. However, in non-equilibrium, time translation invariance does not hold and Green's functions can (and do) depend on $T=(x^{0}+y^{0})/2$ as well. There have been mainly two distinct proposals which attempt to generalize the fluctuation-dissipation theorem to systems not exactly in equilibrium. First, if the system is very near equilibrium, Kadanoff and Baym \cite{kb} propose a generalization of the form
\begin{equation}
C(p, T) = - (1 + 2 f (p_{0},T))\, \text{Im}\, G_{R} (p, T).\label{4}
\end{equation}
Here $f(p_{0},T)$ is a more complicated distribution function that needs to be determined, order by order in perturbation theory, from the self-consistency of \eqref{4}. In some cases, the above ansatz may be extended to non-equilibrium Green's functions as well \cite{czech}.

The second class of out of equilibrium systems to which a generalization has been proposed is known as glassy systems \cite{marinari,ritort,UMB}. This comprises a wide class of systems which have been experimentally studied systematically. Here the relaxation time of the system is much larger than any observation time scale. Therefore, the system approaches equilibrium very slowly such that at any instant of time long after non-equilibrium (but before equilibrium has been attained), the system is in a quasi-equilbrium state. From experimental observations as well as various other considerations, the generalization of the fluctuation-dissipation theorem to glassy systems is proposed to have the form \cite{marinari,ritort,UMB}
\begin{equation}
\frac{1}{\beta_{\rm eff}}\, G_{R} (x^{0},y^{0};\mathbf{p}) = - \theta (x^{0}-y^{0})\,\frac{\partial C(x^{0},y^{0};\mathbf{p})}{\partial y^{0}}.\label{5}
\end{equation}
We note that this relation involves Green's functions in the mixed (time and momentum) space and $\beta_{\rm eff}$ is inversely related to an effective equilibrium temperature during the observation time interval. (The system is in quasi-equilibrium long after non-equilibrium and this effective temperature is a slowly varying function of time.) Furthermore, we note here that while equilibrium Green's functions depend only on $(x^{0}-y^{0})$, this is not true when the system is out of equilibrium. Correspondingly the derivative with respect to the earlier time $y^{0}$ in \eqref{5} is worth emphasizing (namely, in this case $\frac{\partial}{\partial y^{0}} \neq -\frac{\partial}{\partial x^{0}}$). 

In some simpler models, relation \eqref{5} can also be generalized to (we will see this in section {\bf III})
\begin{equation}
\frac{\Omega_{\rm eff}}{2}\,\coth \frac{\beta_{\rm eff}\Omega_{\rm eff}}{2}\, G_{R} (x^{0},y^{0};\mathbf{p}) = - \theta (x^{0}-y^{0})\,\frac{\partial C(x^{0},y^{0};\mathbf{p})}{\partial y^{0}},\label{6}
\end{equation}
which reduces to \eqref{5} in the classical limit. Here $\Omega_{\rm eff}$ is an effective frequency (energy) and both $\Omega_{\rm eff}$ and $\beta_{\rm eff}$ need to be determined, order by order, from the self-consistency of \eqref{6} as well as from other considerations.

Simple soluble models provide an arena for testing these generalized proposals as well as for constructing possible further generalizations. However, there are only a few out of equilibrium models which are soluble. In an earlier paper \cite{frenkel1} (see also \cite{das}), we had studied the large time behavior in one such model. Here we take up this model again to test the validity of these proposals. In section {\bf II}, we describe the model and point out that this behaves like a glassy system in the sense that the system has a long relaxation time. In this section we also give the exact Green's functions for this system. In section {\bf III}, we show that the Green's functions in this model satisfy the generalized glassy equation \eqref{6} and we determine the effective parameters $\Omega_{\rm eff}, \beta_{\rm eff}$. We conclude with a brief summary in section {\bf IV}.

\section{The model and the exact Green's functions}

The model that we study is a simple out of equilibrium model which is soluble. We assume that for negative times, $x^{0}\leq 0$ (the reference time can be arbitrary, we choose it to be zero for simplicity), the model describes a free scalar field theory of mass $m$ in thermal equilibrium at (inverse) temperature $\beta$. A mass correction is introduced for times $x^{0}\geq 0$. Therefore, the Lagrangian density describing the system is given by
\begin{equation}
{\cal L} = \frac{1}{2} \partial_{\mu}\phi \partial^{\mu}\phi - \frac{m^{2}}{2} \phi^{2} - \theta (x^{0}) \frac{\delta m^{2}}{2}\phi^{2} = {\cal L}_{0} + {\cal L}_{I}.\label{7}
\end{equation}
This is a quenched model and the interaction term ${\cal L}_{I} = - \theta(x^{0}) \frac{\delta m^{2}}{2}\phi^{2}$, which  breaks time translation invariance, also takes the theory out of equilibrium. The sudden quench makes this a genuinely out of equilibrium model for any value of the parameter $\delta m^{2}$. As we will see towards the end of section {\bf III}, this model mimics the behavior of a hot scalar field plasma undergoing a sudden expansion (for $\delta m^2  < 0$) or contraction (for $\delta m^2 > 0$), which cools or heats up on time scales much smaller than the relaxation time of the fields \cite{R+W}.

We note here that this simple model captures the essential out of equilibrium features of more realistic self-interacting scalar field theories, where the interaction term has the form $- \lambda\theta(x^{0})\phi^{2n}, n\geq 2$, in the penguin approximation \cite{bedaque}. However, unlike the present model which is soluble, the more realistic models are studied partly analytically and numerically \cite{boyanovsky, berges, berges1}.

The free retarded and correlated Green's functions in this model have the momentum space forms
\begin{align}
G_{R}^{(0)}(p) & = \frac{1}{p^{2}-m^{2} + i\epsilon \text{sgn}(p_{0})},\notag\\
C^{(0)}(p) & = -\frac{1}{2i} G_{c}^{(0)}(p) = \pi \left(1 + 2n (|p_{0}|)\right) \delta(p^{2}-m^{2}),\label{8}
\end{align}
where $\epsilon$ denotes the infinitesimal Feynman regularization parameter which is to be taken to zero, $\epsilon\rightarrow 0^{+}$, only at the end of the calculation. The advanced Green's function can be obtained from \eqref{8} as $G_{A}^{(0)}(p) = G_{R}^{(0)}(-p)$. In the mixed space, the free Green's functions have the forms
\begin{align}
G_{R}^{(0)} (x^{0}-y^{0};\omega) & = -\frac{1}{\omega} \theta(x^{0}-y^{0}) \sin \omega(x^{0}-y^{0})\notag\\
& = G_{A} (y^{0}-x^{0};\omega),\notag\\
C^{(0)} (x^{0}-y^{0};\omega) & = \frac{1}{2\omega} (1+2n(\omega)) \cos \omega(x^{0}-y^{0}),\label{9}
\end{align}
where we have identified $\omega = \sqrt{\mathbf{p}^{2} + m^{2}}$.

Treating ${\cal L}_{I} = - \theta(x^{0})\frac{\delta m^{2}}{2}\phi^{2}$ as a perturbation, the exact Green's functions can now be calculated and have the mixed space forms \cite{frenkel1}
\begin{widetext}
\begin{align}
& G_{R} (x^{0},y^{0};\omega,\Omega) = \theta(x^{0}-y^{0})\left[-\frac{1}{\omega}\theta(-x^{0})\theta(-y^{0}) \sin\omega (x^{0}-y^{0}) -\frac{1}{\Omega}\theta(x^{0})\theta(y^{0}) \sin\Omega (x^{0}-y^{0})\right.\notag\\
&\qquad +\frac{1}{2\omega} \theta(x^{0})\theta(-y^{0}) \left.\left(\left(1-\frac{\omega}{\Omega}\right)\sin(\Omega x^{0} + \omega y^{0}) - \left(1+\frac{\omega}{\Omega}\right)\sin (\Omega x^{0} -\omega y^{0})\right)\right],\label{10a}\\
&C (x^{0},y^{0},\omega,\Omega) = \frac{1}{2\omega} (1+2 n(\omega)) \Big[\theta(-x^{0})\theta(-y^{0}) \cos \omega (x^{0}-y^{0})\notag\\
&\qquad +\frac{1}{2}\theta(-x^{0})\theta(y^{0})\left(\left(1+\frac{\omega}{\Omega}\right)\cos (\omega x^{0} -\Omega y^{0})+\left(1-\frac{\omega}{\Omega}\right) \cos(\omega x^{0} + \Omega y^{0})\right)\notag\\
&\qquad +\frac{1}{2}\theta(x^{0})\theta(-y^{0})\left(\left(1+\frac{\omega}{\Omega}\right)\cos (\Omega x^{0}-\omega y^{0})+\left(1-\frac{\omega}{\Omega}\right) \cos(\Omega x^{0} + \omega y^{0})\right)\notag\\
&\qquad + \frac{1}{2}\theta(x^{0})\theta (y^{0})\left(\frac{(\Omega^{2}+\omega^{2})}{\Omega^{2}}\cos \Omega (x^{0}-y^{0}) + \frac{(\Omega^{2}-\omega^{2})}{\Omega^{2}} \cos \Omega(x^{0}+y^{0})\right)\Big],\label{10}
\end{align}
\end{widetext}
where we have denoted 
\begin{equation}
\Omega^{2} = \omega^{2} + \delta m^{2}.\label{11}
\end{equation}

In writing the exact Green's functions in \eqref{10a} and \eqref{10}, we have suppressed the exponentials involving the regularization parameter for simplicity, but they are quite relevant in understanding the large time behavior of the system. For example, for $x^{0},y^{0} > 0$ (namely, after the quench), the correlated Green's function has the form (with the regularization parameter included)
\begin{widetext}
\begin{equation}
C(x^{0},y^{0};\omega,\Omega) = \frac{1}{4\omega} (1 + 2 n(\omega))\left(\frac{\Omega^{2}+\omega^{2}}{\Omega^{2}} e^{-\epsilon |x^{0}-y^{0}|} \cos \Omega(x^{0}-y^{0}) + \frac{\Omega^{2}-\omega^{2}}{\Omega^{2}} e^{-\epsilon (x^{0}+y^{0})} \cos \Omega(x^{0}+y^{0})\right),\label{12}
\end{equation}
\end{widetext}
and so on. Therefore, the regularization parameter plays the role of the inverse relaxation time which in the present case is large. Correspondingly the system behaves like a glassy system. It is worth pointing out here that in a more realistic model with a genuine scalar interaction, the damping comes from the imaginary part of the self-energy which is mimicked, in this simple theory, by the regularization parameter.

\section{Comparison with the glassy equation}

Our Lagrangian density \eqref{7} describes a quenched, out of equilibrium model which has the characteristics of a glassy system. Therefore, for any value of $\delta m^{2}$, we do not expect it to satisfy the Kadanoff-Baym relation \eqref{4} which is assumed to hold only in systems near equilibrium. (Namely, as we have already pointed out in section {\bf II}, this is a genuinely out of equilibrium model for any value of $\delta m^{2}$ because of the sudden quench.) Fourier transforming the time difference $(x^{0}-y^{0})$ of the Green's functions in \eqref{10a} and \eqref{10}, we have, in fact, explicitly verified that relation \eqref{4} does not hold in the present case. We note that even if we assume $\delta m^{2}$ to be small, even to lowest order in perturbation theory, the corresponding contributions to the Green's functions violate
the Kadanoff-Baym ansatz. For example, to order $\delta m^2$, we get terms of the form $x^0$ $\sin[\omega (x^0 -y^0)]$ which are neither translationally invariant nor slowly varying. In fact, such terms become divergent 
at large times, which reflect the pinch singularities present in thermal perturbation theory \cite{frenkel1}.

Our interest, therefore, is to check whether relation \eqref{5} or \eqref{6} is satisfied in the present soluble model.
To begin with, let us note from \eqref{5} or \eqref{6} that because of the $\theta(x^{0}-y^{0})$ term on the right hand side, we have to consider only the terms in $C(x^{0},y^{0};\omega,\Omega)$ where $x^{0}>y^{0}$. This eliminates the second term in $C(x^{0},y^{0};\omega,\Omega)$ in eq. \eqref{10}. Taking the derivative of this with respect to $y^{0}$ we obtain
\begin{widetext}
\begin{align}
& \theta(x^{0}-y^{0}) \frac{\partial C(x^{0},y^{0};\omega,\Omega)}{\partial y^{0}} = \theta(x^{0}-y^{0})\frac{(1+2n(\omega))}{2\omega} \Big[\theta(-x^{0})\theta(-y^{0})\omega \sin\omega(x^{0}-y^{0}) \notag\\
&\qquad + \frac{\theta(x^{0})\theta(-y^{0})}{2}\Big(\Big(1+\frac{\omega}{\Omega}\Big)\omega\sin(\Omega x^{0}-\omega y^{0}) - \Big(1-\frac{\omega}{\Omega}\Big)\omega\sin (\Omega x^{0}+\omega y^{0})\Big)\notag\\
&\qquad + \frac{\theta(x^{0})\theta(y^{0})}{2}\Big(\frac{\Omega^{2}+\omega^{2}}{\Omega} \sin \Omega(x^{0}-y^{0}) - \frac{\Omega^{2}-\omega^{2}}{\Omega} \sin \Omega (x^{0}+y^{0})\Big)\Big].\label{13}
\end{align}
\end{widetext} 

Let us look at the expression  in \eqref{13} long after the quench where observations on glassy systems are made. In this regime $x^{0},y^{0} \gg 0$ with $x^{0}-y^{0}$ (observation interval) finite so that eq. \eqref{13} takes the form
\begin{widetext}
\begin{equation}
\theta(x^{0}-y^{0})\,\frac{\partial C(x^{0},y^{0};\omega,\Omega)}{\partial y^{0}} = \theta(x^{0}-y^{0})\frac{(1+2n(\omega))}{4\omega} \frac{\Omega^{2}+\omega^{2}}{\Omega} \sin\Omega(x^{0}-y^{0}),\label{14}
\end{equation}
\end{widetext}
where we have used the regularization described in \eqref{12} to set the second term to zero for $x^{0}-y^{0}$ finite but $x^{0}+y^{0}$ large. Comparing this result with the retarded Green's function in \eqref{10a} in the same time regime, we note that we can write
\begin{widetext}
\begin{equation}
-\theta(x^{0}-y^{0})\frac{\partial C}{\partial y^{0}}  = \frac{(\Omega^{2}+\omega^{2})\coth\frac{\beta\omega}{2}}{4\omega}\,G_{R} = \frac{\Omega}{2}\coth \frac{\beta_{\rm eff}\Omega}{2}\,G_{R}.\label{15}
\end{equation}
\end{widetext}
Here the effective temperature can be determined from 
\begin{equation}
\coth \frac{\beta_{\rm eff}\Omega}{2} = \frac{\Omega^{2}+\omega^{2}}{2\omega\Omega} \coth \frac{\beta\omega}{2}.\label{16}
\end{equation}

Therefore, we see that, in this model, the generalized glassy equation \eqref{6} holds which reduces to \eqref{5} in the high temperature (classical) limit. In this case, we note that $\Omega_{\rm eff} = \Omega$. Formally, eq. \eqref{16} is solved for the effective temperature as
\begin{align}
\beta_{\rm eff} & = \frac{2}{\Omega} \text{Arcth}\left[\frac{\Omega^{2}+\omega^{2}}{2\omega\Omega}\coth \frac{\beta\omega}{2}\right]\notag\\
& = \frac{4\omega}{\Omega^{2}+\omega^{2}}\tanh \frac{\beta\omega}{2} + \cdots .\label{17}
\end{align}
When $\beta\omega \ll 1$, this leads to (see \eqref{11})
\begin{equation}
\frac{1}{\beta_{\rm eff}} \simeq \frac{\Omega^{2}+\omega^{2}}{2\omega^{2}}\frac{1}{\beta} = \left(1 + \frac{\delta m^{2}}{2\omega^{2}}\right) \frac{1}{\beta}\label{18}
\end{equation}
where $1/\beta$ is proportional to the temperature. 

This relation can be interpreted as saying that the case $\delta m^{2} > 0$ can be understood as corresponding to a rapid contraction (of a scalar field plasma) leading to a higher effective temperature, while $\delta m^{2} < 0$ can be understood as a rapid expansion leading to a lower effective temperature.

We note that we have a soluble model which yields exact Green's functions. Therefore, in this model we can also explore regimes other than the glassy regime. In particular, we note from \eqref{10a} and \eqref{13} that in the regimes $x^{0},y^{0} < 0$ and $x^{0}>0, y^{0}<0$, the exact Green's functions also satisfy a glassy-like equation \eqref{6}, namely
\begin{equation}
-\theta(x^{0}-y^{0})\frac{\partial C}{\partial y^{0}} = \frac{\omega}{2} \coth \frac{\beta\omega}{2} G_{R}.\label{19}
\end{equation}
In other words, when $y^{0} <0$, the Green's functions satisfy the same simple equation independent of whether $x^{0}>0$ or $x^{0}<0$ (although the forms of the Green's functions in the two regimes are quite different). This emphasizes the special role played by the earlier time $y^{0}$. Furthermore, in both these regimes $\Omega_{\rm eff}=\omega$ and $\beta_{\rm eff}=\beta$. This is easily understood in the regime $x^{0},y^{0}<0$ (initial equilibrium regime), while the behavior in the regime $x^{0}>0, y^{0}<0$ originates from causality which singles out the earlier time.

\section{Conclusion}

We have studied the relation between the correlated and the retarded Green's functions in non-equilibrium, within the context of an exactly soluble quenched model. This model is characterized by a slow approach to equilibrium at large times which is similar to the behavior observed in glassy systems. We have shown that this relation is a generalization of the differential form of the classical fluctuation-dissipation theorem which, in general, involves an effective temperature with a manifest dependence on time sectors. The earlier time $y^{0}$ appearing in the retarded Green's function has a special role in this relation, which is a consequence of causality. When $y^{0} < 0$ (the quench reference time has been chosen to be zero for simplicity), this is similar to the fluctuation-dissipation relation in equilibrium. When $y^{0} > 0$, but small, no such relation holds since in this regime the system is far away from stationarity. On the other hand, for large and positive values of $y^{0}$, the system approaches asymptotically a near-stationary state. In this case, a generalized differential form of the fluctuation-dissipation relation exists which contains an effective temperature. We have determined this temperature which is higher or lower than the initial equilibrium temperature, depending respectively on whether the quench leads to a sudden contraction or expansion of the system.

\bigskip

\noindent{\bf Acknowledgments}
\bigskip

 A. D. would like to thank the Departamento de F\'{i}sica Matem\'{a}tica in USP for hospitality where this work was done. This work was supported in part by USP and by CNPq (Brazil).

\end{document}